\documentclass[apl,twocolumn,showpacs]{revtex4}
\usepackage{graphicx} 

\usepackage{amsmath,amsthm,amssymb,mathrsfs}
\usepackage{bm}

\begin{document}

\title{Theory of spin accumulation and spin transfer torque in a magnetic domain wall}

\author{Tomohiro Taniguchi$^{1,2}$, Jun Sato$^{1}$, and Hiroshi Imamura$^{1}$}
 \affiliation{
 ${}^{1}$ 
 Nanotechnology Research Institute, AIST, Tsukuba, Ibaraki 305-8568, Japan, \\ 
 ${}^{2}$ 
 Institute of Applied Physics, University of Tsukuba, Tsukuba, Ibaraki 305-8573, Japan
 }

 \date{\today} 
 \begin{abstract}
  {
  We studied the spin accumulation and spin transfer torque in a magnetic
  domain wall by solving the Boltzmann equation for spin accumulation
  with the diffusion approximation.  We obtained analytical
  expressions of spin accumulation and spin transfer torque. Both the
  adiabatic and the non-adiabatic components of the spin transfer torque
  oscillate with the thickness of the domain wall.  We showed that the
  oscillating component plays a dominant role in the non-adiabatic
  torque when the domain wall is thinner than the spin-flip length.  We
  also showed that the magnitude of the non-adiabatic torque
  is inversely proportional to the thickness of the domain wall.
  }
 \end{abstract}

 \maketitle



 Spin-dependent electron transport in magnetic nanostructures results in many
 interesting phenomena such as the giant magnetoresistance effect
 \cite{baibich88} and current-induced magnetization dynamics
 \cite{slonczewski96,berger96}.  Recently, such spin-dependent phenomena
 in magnetic domain walls have been investigated, due to great interest in their potential 
 application to spin-electronics devices such as spin-motive-force
 memory \cite{maekawa06,barnes06} and racetrack memory \cite{parkin08}.
 In these devices, data are stored by moving the domain wall 
 using spin transfer torque.

 In 2004, Zhang and Li showed that spin transfer torque in a domain wall can
 be decomposed into two parts, the adiabatic and non-adiabatic torque \cite{zhang04}. 
 The adiabatic torque lies along the spatial gradient of the
 local magnetization while the non-adiabatic torque is perpendicular to 
 this direction. Assuming that the spin accumulation obeys the
 phenomenological diffusion equation and is spatially independent of the
 domain wall, Zhang and Li showed that the ratio of the magnitudes of the
 adiabatic and non-adiabatic torques is determined by the precession
 frequency of the spin accumulation due to the exchange coupling and the
 spin-flip scattering time, and that the non-adiabatic torque is about
 two orders of magnitude smaller than the adiabatic torque.

 The thickness of a domain wall is determined by the competition of the
 exchange coupling between the localized magnetizations and the magnetic
 anisotropy, and is usually on the order of 100 nm for conventional
 ferromagnetic metals such as Fe, Co, Ni, and their alloys.  However, 
 recent developments in the processing technology for nanostructures 
 have allowed the production of a domain wall whose thickness is on the order of 1-10 nm, 
 by reinforcing the shape anisotropy of a magnetic nano-wire
 \cite{ebels00} or trapping a domain wall in a current-confined-path
 geometry \cite{fuke07}. These developments motivated us to study the transport
 phenomena in a thin domain wall \cite{levy97,sato08,matsushita08}.
 For such a thin domain wall, we cannot assume the spatial independence 
 of the spin accumulation.  Thus, it is important to estimate spin transfer
 torque by taking into account the spatial variation in the spin
 accumulation, which would be different from the estimation by 
 Zhang and Li \cite{zhang04}.  Recently, Vanhaverbeke and Viret calculated spin
 transfer torque in a thin domain wall by numerically solving the
 time-dependent Larmor equation of the spin accumulation in a moving frame
 and showed that the non-adiabatic torque is one order of magnitude larger
 than that estimated by Zhang and Li when the thickness of the domain wall
 is comparable to the Lamor precession length \cite{vanhaverbeke07}.


 In this paper, we study spin accumulation and spin transfer torque in a
 domain wall by solving the Boltzmann equation with a diffusion
 approximation. 
 We obtained the analytical expressions of spin accumulation and spin
 transfer torque. 
 Both the adiabatic and the non-adiabatic components of the spin transfer torque
 oscillate with the thickness of the domain wall.  We show that the 
 oscillation plays a dominant role in the non-adiabatic torque
 when the domain wall thickness is less than the spin-flip length, 
 which is defined by the product of the Fermi velocity and the spin-flip scattering time.
 For a domain wall that is much thinner than the spin-flip length,
 the non-adiabatic torque is about one order of magnitude smaller than the
 adiabatic torque, which is one order of magnitude larger than that
 estimated by Zhang and Li \cite{zhang04} and qualitatively consistent
 with the results of Vanhaverbeke and Viret \cite{vanhaverbeke07}.  
 We also showed that the magnitude of the non-adiabatic torque
 is inversely proportional to the thickness of the domain wall.


 We considered electron transport in a one-dimensional magnetic nanowire with
 a 180${}^{\circ}$ domain wall which lies over $-d/2\!\le\!x\!\le\!d/2$, where
 $d$ is the thickness of the domain wall.  We assume that the interaction
 between the conducting (s-like) electrons and the localized (d-like)
 electrons is described by an sd exchange interaction, 
 $\hat{H}_{\rm sd}\!=\!-(J/2)\hat{\bm{\sigma}}\cdot\hat{\mathbf{S}}$, where
 $\hat{\bm{\sigma}}$ is the vector of the Pauli matrices, $J$ is the sd
 exchange coupling constant and
 $\hat{\mathbf{S}}(x)\!=\!(0,-\sin\theta,\cos\theta)$ is the unit vector
 pointing along the direction of the localized spin angular momentum.
 The angle $\theta(x)$ is given by 0 for $x\!<\!-d/2$, $(\pi/d)(x+d/2)$
 for $-d/2\!<\!x\!<\!d/2$, and $\pi$ for $x\!>\!d/2$, respectively. 

 Following $\breve{{\rm S}}$im\'anek and Rebei
 \cite{simanek01,simanek05}, we employ the rotating frame where basic
 unit vectors are defined as
 $\mathbf{e}_{x}=\alpha^{-1}\hat{\mathbf{S}}\times(\partial\hat{\mathbf{S}}/\partial
 x)$, $\mathbf{e}_{y}=-\alpha^{-1}\partial\hat{\mathbf{S}}/\partial x$
 and $\mathbf{e}_{z}=\hat{\mathbf{S}}$, respectively.  
 We assume that the direction of the localized spin varies slowly compared
 to the Fermi wavelength
 $\lambda_{\rm F}$, i.e., $\alpha={\rm d}\theta/{\rm d}x \ll
 2\pi/\lambda_{\rm F}$; thus, we could neglect the higher-order terms of $\alpha$
 in the following calculation.
 The spin accumulation and spin transfer torque in a domain wall are
 obtained by solving the Boltzmann equation for the Wigner function
 defined as
 $\hat{f}(x,p_{x})=[f(x,p_{x})\hat{1}+\mathbf{g}(x,p_{x})\cdot\hat{\bm{\sigma}}]/2$,
 where $f(x,p_{x})$ and $\mathbf{g}(x,p_{x})$ represent the charge and
 spin distribution functions, respectively. The spin accumulation $\mathbf{s}$ and the
 spin current density $\mathbf{j}$ are defined as 
 \begin{align}
  &
  \mathbf{s}
  =
  \int 
  \mathbf{g}
  \frac{{\rm d}^{3}\mathbf{p}}{(2\pi\hbar)^{3}}\ ,
  \\
  &
  \mathbf{j}
  =
  \int
  v_{x}\mathbf{g} 
  \frac{{\rm d}^{3}\mathbf{p}}{(2\pi\hbar)^{3}}\ ,
 \end{align}
respectively.
 It should be noted that the dimensions of $\mathbf{s}$ and
 $\mathbf{j}$ are density and density times velocity, respectively.
 The diffusion approximation, $\int v_{x}^{2}\mathbf{g}{\rm
 d}^{3}\mathbf{p}/(2\pi\hbar)^{3} \simeq (v_{\rm F}^{2}/3)\mathbf{s}$, is
 applied to the Boltzmann equation \cite{simanek01}. 
 Up to the first order of $\alpha$, the transverse components of the
 spin accumulation, $s_{x}$ and $s_{y}$, and spin current, $j_{x}$ and
 $j_{y}$, obey the following equations; 
 \begin{align}
  &\frac{\partial s_{x}}{\partial x} 
  =
  -\frac{1}{2D} j_{x} 
  +
  \frac{\omega_{J}\tilde{T}}{D} j_{y}\ ,
  \label{eq:partial_sx}
  \\
  &
  \frac{\partial s_{y}}{\partial x} 
  =
  -\frac{\omega_{J}\tilde{T}}{D} j_{x} 
  -
  \frac{1}{2D} j_{y}\ ,
  \label{eq:partial_sy}
  \\
  &
  \frac{\partial j_{x}}{\partial x}
  -
  \omega_{J}s_{y}
  +
  \frac{2}{\tau_{\rm sf}} s_{x}
  =
  0\ ,
  \label{eq:partial_jx} 
  \\
  &
  \frac{\partial j_{y}}{\partial x} 
  +
  \omega_{J}s_{x} 
  +
  \frac{2}{\tau_{\rm sf}} s_{y}
  =
  \alpha j_{z}\ ,
  \label{eq:partial_jy}
 \end{align}
 where $\omega_{J}=J/\hbar$ is the Larmor precession frequency, 
 $\tilde{T}$ is the momentum relaxation time,
 $\tau_{\rm sf}$ is the spin-flip scattering time and $D=v_{\rm
 F}^{2}\tilde{T}/3$ is the diffusion constant \cite{simanek01,simanek05}.
 The longitudinal spin current $j_{z}$ in Eq. \eqref{eq:partial_jy} is
 given by $j_{z}=\beta
 j_{e}/(-e)$, where $\beta$ and $j_{e}$ are the spin polarization factor
 and the electric current density, respectively.  
 In our definition, the positive electric current corresponds to 
 the electron flow along the $-x$ direction.

 The physics behind  Eqs. (\ref{eq:partial_sx})-(\ref{eq:partial_jy})
 are as follows. 
 Traveling through the domain wall, the conducting electrons vary the
 direction of their spin along the localized spin angular momentum
 $\hat{\mathbf{S}}$.  Then, spin accumulation and spin current
 polarized along the $y$-direction ($\propto
 \partial\hat{\mathbf{S}}/\partial x$) are induced; see
 Eq. (\ref{eq:partial_jy}).  The accumulated spins precess around
 $\hat{\mathbf{S}}$ due to the sd exchange coupling with the precession
 frequency $\omega_{J}$.  Then, the $x$-components of the spin
 accumulation and spin current are induced [see
 Eq. (\ref{eq:partial_jx})].  Equations (\ref{eq:partial_sx}) and
 (\ref{eq:partial_sy}) relate the spin accumulation and spin current by
 the diffusion constant.


 Before estimating the spin accumulation and spin transfer torque in a
 domain wall, we should emphasize the validity of our calculations.  Since
 Eqs. (\ref{eq:partial_sx})-(\ref{eq:partial_jy}) are obtained by
 applying the diffusion approximation to the Boltzmann equation, they are
 valid for $d \ge l_{\rm mfp}$, where $l_{\rm mfp}=v_{\rm F}\tilde{T}$ is
 the mean-free-path of the conducting electrons.  For a domain wall 
 the thickness of which is much smaller than the mean free path, i.e.,
 $d \ll l_{\rm mfp}$, the Boltzmann equation should be solved 
 without the diffusion approximation.  Moreover, in such a very thin
 domain wall, we cannot neglect the higher-order terms of $\alpha$.


 We assume that the transverse spin accumulation and spin current are such 
 that they vanish at the limit of $|x|\to\infty$ and are continuous at $x=\pm
 d/2$.  Then, solving Eqs. (\ref{eq:partial_sx})-(\ref{eq:partial_jy}),
 the transverse spin accumulations in the domain wall are obtained as
 $s_{x}={\rm Re}[s_{+}]$ and $s_{y}={\rm Im}[s_{+}]$, where
 \begin{equation}
  s_{+} 
   =
   \frac{\pi(1+{\rm i}\zeta)j_{z}}{\omega_{J}d(1+\zeta^{2})}
   \left[
    1
    -
    \exp
    \left(
     -\frac{d}{2\ell}
    \right)
    \cosh
    \left(
     \frac{x}{\ell}
    \right)
   \right]\ .
   \label{eq:s_plus}
 \end{equation}
Here $\zeta=2/(\omega_{J}\tau_{\rm sf})$ and $\ell$ is given by 
 \begin{equation}
  \frac{1}{\ell} 
   =
   \sqrt{
   \frac{1}{2D}
   \left(
    1+2{\rm i}\omega_{J}\tilde{T}
   \right)
   \left(
    {\rm i}\omega_{J}
    +
    \frac{2}{\tau_{\rm sf}} 
   \right)
   }\ ,
 \end{equation}
where $k_{\rm r}={\rm Re}[1/\ell]$ and $k_{\rm i}={\rm Im}[1/\ell]$
 characterize the oscillation and damping of $s_{x}$ and $s_{y}$ due to
 the sd exchange coupling and the spin-dependent scattering,
 respectively.  As shown in Eq. (\ref{eq:s_plus}), the transverse spin
 accumulations can be decomposed into spatially independent (first) and
 dependent (second) parts.


 \begin{figure}
  \centerline{\includegraphics[width=1.0\columnwidth]{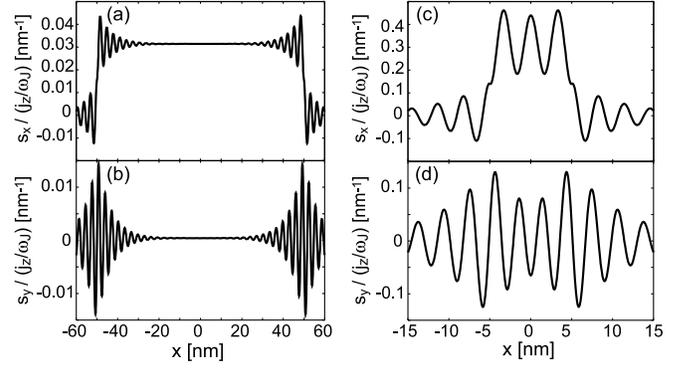}}
  \caption{
  The spatial variation of the transverse spin accumulations
  for a thick ($d=100$ nm) and thin ($d=10$ nm) domain wall; 
  (a) $s_{x}$ for $d=100$ nm, (b) $s_{y}$ for $d=100$ nm, (c) $s_{x}$ for $d=10$ nm and (d) $s_{y}$ for $d=10$ nm, 
  respectively. 
  The magnitudes of $s_{x}$ and $s_{y}$ are divided by $j_{z}/\omega_{J}$, 
  see Eq. (\ref{eq:s_plus}). 
  }
  \vspace{-3ex}
  \label{fig:fig1}
 \end{figure}

 Figure \ref{fig:fig1} shows the spatial dependence of the transverse
 spin accumulations, $s_{x}$ and $s_{y}$, for thick ($d=100$ nm) and
 thin ($d=10$ nm) domain walls, respectively.  For convenience, $s_{x}$ and
 $s_{y}$ are divided by $j_{z}/\omega_{J}$ [see Eq. (\ref{eq:s_plus})].
 The parameters are taken to be $J=1.0$ eV, $\tau_{\rm sf}=10^{-4}$ ns,
 and $l_{\rm mfp}=3.0$ nm, respectively.  The Fermi velocity is given by
 $v_{\rm F}=\sqrt{2\varepsilon_{\rm F}/m}$, where the Fermi energy is
 taken to be $\varepsilon_{\rm F}=5.0$ eV.  These are typical values for
 the conventional transition ferromagnetic metals \cite{gurney93}. As
 shown in Figs. \ref{fig:fig1} (a) and \ref{fig:fig1} (b), for a thick domain wall, the
 spin accumulation in the domain wall is nearly spatially independent
 except at the boundaries of the domain wall $x=\pm d/2$.  On the other
 hand, as shown in Figs. \ref{fig:fig1} (c) and \ref{fig:fig1} (d), for a thin domain
 wall, the spin accumulations vary in the domain wall, and we cannot
 assume the spatial independence of the spin accumulations.


 Let us estimate spin transfer torque in the domain wall, 
 which is defined as 
 \begin{equation}
  \bm{\tau} 
   =
   \int_{-d/2}^{d/2} 
   \omega_{J} 
   \mathbf{s}\times\hat{\mathbf{S}}
   {\rm d}x\ .
 \end{equation}
 The Landau-Lifshitz-Gilbert equation for 
 the localized magnetization $\hat{\mathbf{M}}=-\hat{\mathbf{S}}$ 
 with the torque $\bm{\tau}$ is given by 
 \begin{equation}
  \begin{split}
   \frac{\partial\hat{\mathbf{M}}}{\partial t} 
   =&
   -\gamma\hat{\mathbf{M}}\times\mathbf{B} 
   +
   \alpha_{0}
   \hat{\mathbf{M}}\times\frac{\partial\hat{\mathbf{M}}}{\partial t} 
   \\
   &
   +
   \frac{\gamma\hbar}{2\pi M}
   \tau_{y}
   \frac{\partial\hat{\mathbf{M}}}{\partial x}
   +
   \frac{\gamma\hbar}{2\pi M} 
   \tau_{x}
   \hat{\mathbf{M}}\times\frac{\partial\hat{\mathbf{M}}}{\partial x}\ ,
   \label{eq:LLG}
  \end{split}
 \end{equation}
 where $\gamma$ is the gyromagnetic ratio, $\mathbf{B}$ is the effective
 magnetic field, $M$ is the magnitude of the magnetization 
 and $\alpha_{0}$ is the Gilbert damping
 constant.  $\tau_{y}=\mathbf{e}_{y}\cdot\bm{\tau}$ and
 $\tau_{x}=\mathbf{e}_{x}\cdot\bm{\tau}$ correspond to the adiabatic and
 non-adiabatic torque, respectively.  By using Eq. (\ref{eq:s_plus}), we
 find that
 \begin{equation}
  \begin{split}
   \tau_{y} 
   =&
   \frac{\pi\beta j_{e}}{e(1+\zeta^{2})} 
   \\
   &-
   \frac{\pi\beta j_{e}[k_{\rm r}-{\rm e}^{-k_{\rm r}d}(k_{\rm r}\cos k_{\rm i}d-k_{\rm i}\sin k_{\rm i}d)]}
   {ed(1+\zeta^{2})(k_{\rm r}^{2}+k_{\rm i}^{2})}
   \\
   &-
   \frac{\pi\zeta\beta j_{e}[k_{\rm i}-{\rm e}^{-k_{\rm r}d}(k_{\rm i}\cos k_{\rm i}d+k_{\rm r}\sin k_{\rm i}d)]}
   {ed(1+\zeta^{2})(k_{\rm r}^{2}+k_{\rm i}^{2})}\ ,
   \label{eq:tau_y}
  \end{split}
 \end{equation}
 \begin{equation}
  \begin{split}
   \tau_{x} 
   =&
   -\frac{\pi\zeta\beta j_{e}}{e(1+\zeta^{2})} 
   \\
   &-
   \frac{\pi\beta j_{e}[k_{\rm i}-{\rm e}^{-k_{\rm r}d}(k_{\rm i}\cos k_{\rm i}d+k_{\rm r}\sin k_{\rm i}d)]}
   {ed(1+\zeta^{2})(k_{\rm r}^{2}+k_{\rm i}^{2})}
   \\
   &+
   \frac{\pi\zeta\beta j_{e}[k_{\rm r}-{\rm e}^{-k_{\rm r}d}(k_{\rm r}\cos k_{\rm i}d-k_{\rm i}\sin k_{\rm i}d)]}
   {ed(1+\zeta^{2})(k_{\rm r}^{2}+k_{\rm i}^{2})}\ .
   \label{eq:tau_x}
  \end{split}
 \end{equation}
 The first terms of Eqs. (\ref{eq:tau_y}) and (\ref{eq:tau_x}) are
 identical to the adiabatic and non-adiabatic torque estimated by Zhang
 and Li \cite{zhang04}, respectively.  These first terms arise from the
 spatially independent part of the spin accumulation, i.e., the first term of
 Eq. (\ref{eq:s_plus}).  It should be noted that these terms are
 independent of the thickness of the domain wall $d$.  For a thick domain
 wall, these first terms are dominant for spin transfer torque, and the
 ratio of the magnitude of the adiabatic and non-adiabatic torque,
 $|\tau_{x}/\tau_{y}|$, is given by $\zeta\simeq 10^{-2}$ \cite{zhang04}.
 On the other hand, the second and third terms of Eqs. (\ref{eq:tau_y})
 and (\ref{eq:tau_x}) arise from the spatial variation in the spin
 accumulation, i.e., the second term of Eq. (\ref{eq:s_plus}).  As shown
 in Figs. \ref{fig:fig1} (c) and \ref{fig:fig1} (d), for a thin domain wall, we cannot
 neglect the spatial variation in the transverse spin accumulation, and
 these second and third terms dominate the spin transfer torque.  It
 should be noted that these terms are inversely proportional to the
 thickness $d$.  Thus, for a thin domain wall, the strength of the spin
 transfer torque is considerably different from that estimated
 by Zhang and Li \cite{zhang04}.


 \begin{figure}
  \centerline{\includegraphics[width=0.8\columnwidth]{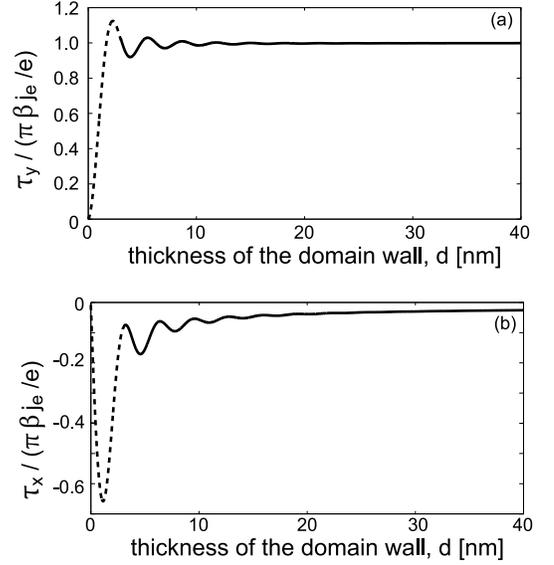}}
  \caption{ (a) The strength of the adiabatic torque $\tau_{y}$
  renormalized by $\pi\beta j_{e}/e$ against the thickness of the domain
  wall is shown.  (b) The strength of the non-adiabatic torque
  $\tau_{x}$ renormalized by $\pi\beta j_{e}/e$ against the thickness of
  the domain wall is shown.  For $d\le l_{\rm mfp}=3$ nm, the
  diffusion approximation cannot be applied to the Boltzmann equation,
  and thus, torque below $d\le l_{\rm mfp}$ denoted by the dotted line
  is not valid. }
  \vspace{-3ex}
  \label{fig:fig2}
 \end{figure}


 Figure \ref{fig:fig2} shows the strength of the adiabatic torque
 $\tau_{y}$ and the non-adiabatic torque $\tau_{x}$ renormalized by
 $\pi\beta j_{e}/e$ against the thickness of the domain wall $d$.  We
 denote the torque for $d \le l_{\rm mfp}$ by the dotted line because our
 calculations are restricted for $d\ge l_{\rm mfp}$; thus, the
 torques for $d \le l_{\rm mfp}$ are not valid.  As shown in
 Fig. \ref{fig:fig2}, for $d \geq 30$ nm, spin transfer torque is nearly
 independent of the thickness $d$.  On the other hand, for $d \ll 30$ nm,
 the strength of the non-adiabatic torque increases as the 
 thickness $d$ decreases.  For $d \leq 10$ nm, $|\tau_{x}/\tau_{y}|\simeq 10^{-1}$,
 which is one order of magnitude larger than that estimated by Zhang and Li
 \cite{zhang04}.  Moreover, for $d\leq 10$ nm, the spin transfer torque
 oscillates against the thickness $d$ with the period of the oscillation
 given by $2\pi/k_{\rm i}$.


 Let us reveal the parameters which characterize the above behavior of the
 spin transfer torque.  Assuming that $k_{\rm r}\!\simeq\! \sqrt{3}/(4l_{\rm
 mfp})\!\ll\! k_{\rm i}\!\simeq\! \sqrt{3}\omega_{J}/v_{\rm F}$ (Ref. \cite{simanek05})
 and $d\!\gg\! l_{\rm mfp}$, we find that $\tau_{y}\!\simeq\! \pi\beta j_{e}/e$
 and $\tau_{x}\!\simeq\! -\pi\zeta\beta j_{e}/e - \pi\beta j_{e}/(edk_{\rm
 i})$, respectively.  Thus, for $d\!\gg\!l_{\rm mfp}$, the adiabatic torque
 is nearly independent of the thickness.  On the other hand, for $d
 \!\ll\! 1/(\zeta k_{\rm i})\!\simeq\! l_{\rm sf}/(2\sqrt{3})\!\simeq\!
 40$ nm, where $l_{\rm sf}\!=\!v_{\rm F}\tau_{\rm sf}$ is the spin-flip length, 
 the torque due to the spatial variation in the spin accumulation
 is dominant for the non-adiabatic torque.  For a thin domain wall, the
 ratio of the adiabatic and non-adiabatic torque is characterized by
 $v_{\rm F}/(\sqrt{3}\omega_{J}d)$, which is the ratio of the precession
 frequency of the electrons' spin around the localized spin angular
 momentum due to the sd exchange coupling and the angular velocity of the
 rotation of the exchange field in the domain wall.  For $d\!=\!10$ nm,
 $v_{\rm F}/(\sqrt{3}\omega_{J}d)\simeq 10^{-1}$.  The oscillation period
 is given by $2\pi/k_{\rm i}\!\simeq\! 2\pi v_{\rm
 F}/(\sqrt{3}\omega_{J})\!\simeq\! 2.5$ nm.  These estimations can be confirmed by the plots
 shown in Fig. \ref{fig:fig2}.


 When the precession frequency of the electrons' spin around the exchange
 field, $\omega_{J}$, is comparable to the angular velocity of the
 rotation of the exchange field in space, $\pi v_{\rm F}/d$, the direction of the 
 electrons' spin cannot vary their direction adiabatically, and the
 non-adiabaticity, which is sometimes called the mistracking effect,
 plays an important role in the spin-dependent transport phenomena.  For
 example, the terminal velocity of the domain wall motion is proportional
 to the ratio of the adiabatic and non-adiabatic torque
 \cite{thiaville05}. The non-adiabaticity is characterized by a
 dimensionless parameter $\xi\!=\!\pi v_{\rm F}/(2d\omega_{J})$
 \cite{marrows05}. As shown above, the ratio of the adiabatic and
 non-adiabatic torques for a thin domain wall, $v_{\rm
 F}/(\sqrt{d}\omega_{J}d)$, is the first order of $\xi$, while the
 magnetoresistance due to the mistracking effect or spin accumulation
 is on the second order of $\xi$ \cite{levy97,simanek01,simanek05}.  For
 conventional ferromagnetic metals with $d \ge l_{\rm mfp}$, $\xi$ is
 less than unity.  Thus, the non-adiabaticity plays an important role in
 the dynamics of the localized magnetization compared to the
 magnetoresistance.  It should be noted that for a thick domain wall the
 non-adiabatic torque is characterized by $\zeta\!=\!2/(\omega_{J}\tau_{\rm
 sf})$, not $\xi$, as shown by Zhang and Li \cite{zhang04}.


 We compare our results with those of Vanhaverbeke and Viret
 \cite{vanhaverbeke07}.  In Ref. \cite{vanhaverbeke07}, a time-dependent
 phenomenological Larmor equation for the magnetic moment in a
 moving frame is solved numerically, and showed that the non-adiabatic
 torque is shown to be one order of magnitude larger than that estimated by Zhang and Li
 \cite{zhang04} when the thickness of the domain wall is comparable to
 the Larmor precession length $\lambda_{\rm L}\!=\!v_{\rm
 F}/(2\pi\omega_{J})$, which is on the order of a few nanometers.  On the
 other hand, we consider the spin diffusion in the domain wall in a
 steady state by solving the Boltzmann equation in the rotated frame, and
 analytical expressions of the spin accumulation and spin transfer
 torque are obtained.  We show that the strength of the non-adiabatic
 torque increases as the thickness of the domain wall decreases 
 for $d\!\leq\! l_{\rm sf}/(2\sqrt{3})$. Note that the condition is determined
 by the spin-flip length $l_{\rm sf}$ instead of the Larmor precession 
 length $\lambda_{\rm L}$. 
 We also find that the strength of 
 the non-adiabatic torque is characterized by the first order of 
 the non-adiabatic parameter $\xi\!\propto\!1/d$. 
 The non-adiabatic torque does not change its sign, as shown in Fig. 5 in
 Ref. \cite{vanhaverbeke07}.


 Since the diffusion approximation is applied to the Boltzmann equation, 
 the present theory is not applicable to the ballistic region $d \le l_{\rm mfp}$. 
 The spin transfer torque in the ballistic region is obtained by Waintal
 and Vilet \cite{waintal04}.  
 One can easily confirm that our Eq. (\ref{eq:s_plus}) reduces
 to Eq. (12) of Ref. \cite{waintal04} in the limit of $\tilde{T},\tau_{\rm sf} \to \infty$, 
 where they assume that the local spin transfer torque 
 is proportional to the spin accumulation. 
 One might expect a simple connection formula between ballistic and diffusive spin
 transfer torque like Wexler's formula for conductance \cite{wexler66}. 
 However, this is beyond the scope of the present paper.



 In conclusion, we studied spin transfer torque in a domain wall by
 solving the Boltzmann equation for spin accumulation, and found their
 analytical expressions. For a thin domain wall whose thickness is 
 much thinner than the spin-flip length, 
 the ratio of the magnitude of the adiabatic and non-adiabatic
 torque is about $10^{-1}$, which is one order of magnitude larger than that
 estimated in Ref. \cite{zhang04} and consistent with that in 
 Ref. \cite{vanhaverbeke07}.  We also found that the strength of the
 non-adiabatic torque is inversely proportional to the thickness of
 the domain wall.


 The authors would like to acknowledge the valuable discussions they had
 with P. M. Levy, J. Ieda, H. Sugishita, K. Matsushita, and N. Yokoshi. 
 This work was supported by JSPS and NEDO.



\end{document}